\documentclass{aa}
\usepackage{natbib}
\usepackage{graphicx}
\newcommand{\Al}{$^{26}$Al\ }
\newcommand{\about}{$\simeq$}

\begin{document}

\title{SPI Measurements of Galactic \Al}

\author{R. Diehl \inst{1}  
 \and 
 J. Kn\"odlseder \inst{2}
 \and
 G.G. Lichti \inst{1}
 \and
 K. Kretschmer \inst{1}
  \and
  S. Schanne \inst{3}
  \and
  V. Sch\"onfelder \inst{1}
 \and 
 A.W.Strong \inst{1}
 \and
 A.von Kienlin \inst{1}
 \and 
 G. Weidenspointner \inst{2,4,5}
 \and 
 C. Winkler \inst{6}
 \and
 C. Wunderer \inst{1}
}
\institute{
Max-Planck-Institut f\"ur extraterrestrische Physik,
              D-85741 Garching, Germany 
 \and
 Centre d'Etude Spatiale des Rayonnements, 31028 Toulouse, France
 \and
 DSM/DAPNIA/Service d'Astrophysique, CEA Saclay, 91191 Gif-Sur-Yvette, France	
 \and
 NASA/Goddard Space Flight Center, Greenbelt, MD 20771, USA	
 \and
Universities Space Research Association, Seabrook, MD 20706, USA	
 \and
ESA/ESTEC, Science Operations and Data Systems Division (SCI-SD)  
2201 AZ Noordwijk, The Netherlands 		  }

\offprints{R. Diehl}
\mail{rod@mpe.mpg.de}
\date{Received 15 Jul 2003; revised 31 Jul 2003; accepted 02 Sep 2003}
\authorrunning{Diehl et al.}
\titlerunning{SPI Measurements of Galactic \Al}

\abstract { 
The precision measurement of the 1809 keV gamma-ray line
from Galactic $^{26}$Al is one of the goals of the 
SPI spectrometer on INTEGRAL with its Ge detector 
camera. We aim for determination of the detailed shape of this 
gamma-ray line, and its variation for different
source regions along the plane of the Galaxy. 
Data from the first part of the
core program observations of the first mission year have been
inspected. A clear detection of the \Al line 
at \about~5--7~$\sigma$ significance 
demonstrates that SPI will deepen \Al studies.
The line intensity is consistent with expectations from previous
experiments, and the line appears narrower than the 5.4~keV FWHM 
reported by GRIS, more consistent with RHESSI's recent value. 
Only preliminary statements can
be made at this time, however, due to the multi-component background
underlying the signal at \about~40~times higher intensity than 
the signal from Galactic $^{26}$Al.
}

\maketitle

\section{Introduction}
The measurement of 1809 keV emission from Galactic \Al has been one 
of the design goals of the INTEGRAL mission \citep{wink03,herm02}. 
\Al gamma-rays were
discovered already in 1982 by HEAO-C \citep{maho82}, and since 
then are considered direct
proof of ongoing nucleosynthesis in the Galaxy. Several follow-up
experiments have set out to measure details about \Al sources. 

Much has been learned through the Compton Gamma-Ray Observatory, in particular 
with COMPTEL's sky survey over 9 years, which resulted
in an all-sky image in the 1809 keV gamma-ray line 
\citep{plue01,knoed_img99,ober97,dieh95}. 
 This image clearly
demonstrates that emission along the plane of the Galaxy dominates, hence
\Al nucleosynthesis is common throughout the Galaxy, rather than a local
phenomenon of the solar system. The irregular structure of the emission
and alignments of emission maxima with spiral-arm tangents suggested that
massive stars dominate \Al nucleosynthesis \citep{chen95,pran96}. 
This could be 
further substantiated through comparisons with candidate source tracers, and 
through modelling of \Al emission from the Galaxy and specific source regions
based on knowledge about the massive-star populations 
\citep{knoe_mod99,knoed_phd99}. 

The high spectral resolution of Ge detectors of 3~keV (FWHM) at the
\Al line energy of 1808.7~keV is expected to reveal more information about
the sources and their location through Doppler broadenings and shifts, from
Galactic rotation \citep{gehr96} and from dynamics of the \Al gas ejected into the 
interstellar medium. In particular after the GRIS balloon experiment and their
report of a significantly-broadened line \citep{naya96}, 
alternative measurements of the \Al line 
shape were of great interest. GRIS's value 
translates into an intrinsic line width of 5.4~keV, equivalent to a Doppler
broadening of 540~km~s$^{-1}$. Considering the $1.04 \times 10^6$~y decay time of \Al
such a large line width is hard to understand \citep{chen97,stur99}.
Other high-resolution measurements
are in mild conflict with the GRIS result. 
The original HEAO-C measurement was interpreted
as an intrinsically-narrow line of width less than 2 keV \citep{maho84}, 
and the recent RHESSI measurement shows some broadening, however at an
intermediate value of about $2.0 (+0.78, -1.21)$~keV FWHM \citep{smit03}, 
significantly less than the GRIS value.  
SPI on INTEGRAL \citep{vedr03,roqu03} with its competitive spectral resolution 
and the INTEGRAL core observing program \citep{wink03} 
emphasizing exposures of the inner region of the Galaxy
is expected to clarify these questions through high-quality data.
The initial calibration phase early in the mission, which used exposures of
sources in the Cygnus region, had demonstrated an excellent
performance of the instrument and even shown convincing evidence of detections
of diffuse \Al emission from the Cygnus region, supporting these prospects.
In this paper, we report initial analyses of the first part of the inner
Galaxy region survey of the INTEGRAL core program. 

\begin{figure}[ht]
 \includegraphics[width=0.5\textwidth]{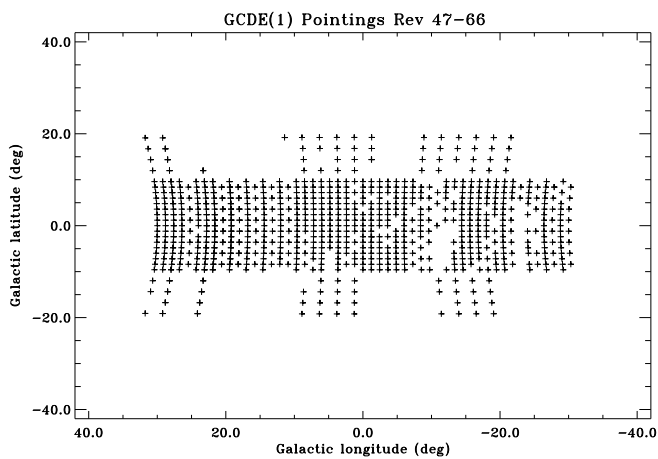}
   \caption{Pointings of the first part of the 
    Galactic-Central-Deep-Exposure analyzed here}
  \label{exposure}
\end{figure}

\section{Observations, Data, and Analysis}
The first year of the INTEGRAL mission will emphasize a deep survey
of the inner part of the Galaxy, devoting about 4 Msec of exposure 
to this region, in a dedicated observing pattern characterized by 
a 2$^\circ$ pitch, extending $\pm$~30$^\circ$ in longitude and
$\pm$~20$^\circ$ in latitude away from the Galactic Center \citep{wink03}. 
Data sharing agreements within the INTEGRAL Science Working Team
imply that results on Galactic \Al for the separate Galactic quadrants
will not be addressed here, but reported in future papers. 
The data used for this initial analysis encompass
INTEGRAL orbits 46--66 (mission days 1157--1216). Not all our analyses use the
exactly same data, different selections were applied from early stages
of processing; a minimum set comprises in total 839 pointings with 0.971~Ms
integrated exposure livetime (see Figure \ref{exposure}).

Energy calibration during this time was derived by fitting instrumental 
background lines at energies 439, 585, 883, 1014, 1369, and 1779~keV as 
accumulated
for each orbit. Calibrated single-detector events 
(``SE'', triggering one single of the 19 Ge detectors), 
and the composite of single and multiple
detector hits (SE+``multiple events, ME'') were analyzed. Note that at the 
\Al line energy 40\% of the measured events are multiples. A systematic
uncertainty of our energy calibration at energies below \about~150~keV
may however distort the energies assigned to multiple events. We therefore
perform independent analyses per event type (SE, SE+ME), to check for such systematics
which could lead to artifacts in the wings of spectral lines. 

\begin{figure}[ht]
 \includegraphics[width=0.5\textwidth]{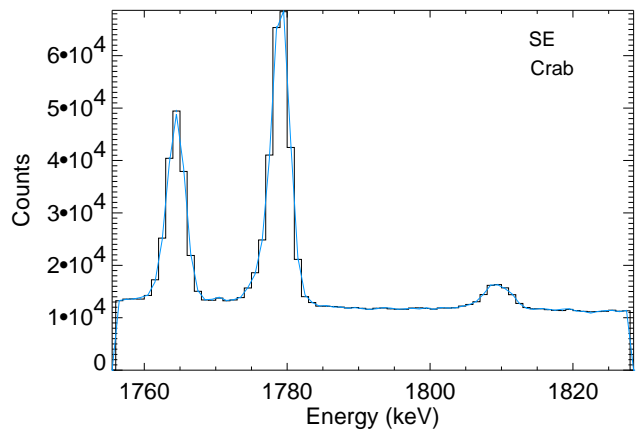}
   \caption{Raw spectrum of Crab reference observations (SE).
   Background dominates the signal, with prominent instrumental lines
   at 1779~keV originating from $^{28}$Al, at 1764~keV from $^{205}$Bi,
   and  at 1809-1811~keV from a blend of $^{26/27}$Na and $^{56}$Mn.
   }
  \label{rawspec_crab}
\end{figure}

\begin{figure}[ht]
 \includegraphics[width=0.5\textwidth]{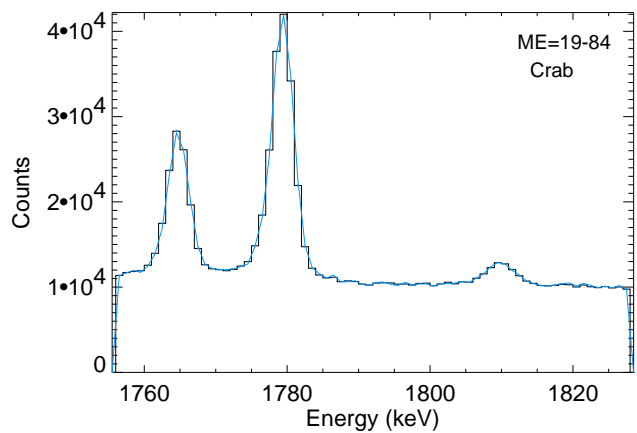}
   \caption{Raw spectrum of Crab reference observations for multiple events (ME).
   The same instrumental lines are seen, slightly weaker if compared to
   continuum background.
   }
  \label{rawspec_crab_ME}
\end{figure}

\begin{figure}[ht]
 \includegraphics[width=0.5\textwidth]{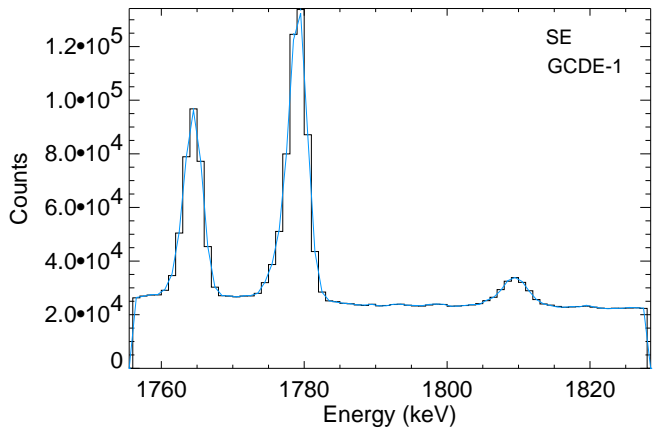}
   \caption{Raw spectrum of GCDE measurements (SE). 
   The form of the instrumental-background spectrum
   is virtually identical to the Crab reference (Fig. \ref{rawspec_crab}),  
   the signal expected from celestial \Al is at the percent level.}
  \label{rawspec_gcde}
\end{figure}

Background dominates the overall signal, 
(see Figures \ref{rawspec_crab}--\ref{rawspec_gcde}), 
so that in the \Al line region
(1809$\pm$4~keV) we measure $5 \times 10^5$ counts from the inner Galaxy, 
while our background reference is 
based on exposures of the Crab region providing 
$1.3 \times 10^5$ counts in the line-region. 
Continuum dominates, but \about~17\% of the total signal is in 
a rather broad line-like feature around 1810~keV. 
This is mostly instrumental background, which needs to be
understood before extracting the \Al signal: For the GCDE, we expect from
COMPTEL measurements a signal strength of about 13000~counts or 2.7\% of
the total measured counts in the line region. Background is expected at
1808.7 keV from excited $^{26}$Mg produced from spallation of Al and
from $\alpha$-captures on Na, and at 1810.7~keV from 
$^{56}$Mn($\beta^-$)$^{56}$Co($EC$)$^{56}$Fe, but other nuclear lines
may contribute \citep{weid03}. 
Correlation analysis with other line features and 
cosmic-ray activation monitors is underway to model details of this
underlying background, both in shape and intensity \citep[e.g.][]{jean03}.  

\begin{figure}[ht]
 \includegraphics[width=0.5\textwidth]{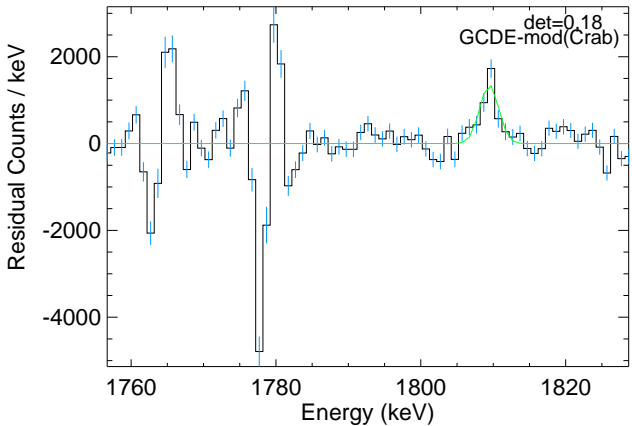}
   \caption{Residual spectrum of GCDE - Crab (SE), after normalization on the
   intensities of the 1764 and 1779 keV instrumental background lines, adjusting
   for detector degradation between the two observations. The residual excess
   around 1809~keV appears above the properly-scaled residuals from the 
   instrumental background lines, which remain from imperfections in the
   subtraction process. Here the coded-mask imaging information is not used.}
  \label{spec_gcde-crab}
\end{figure}

Data analysis is complicated due to the large number (839) of individual
pointings, different measurement times of source and background,
and detector and background evolution within and between them.
For a strong source, one may hope to subtract a sufficiently well-defined
background and then see a source signal. Indeed, when we use exposures
from the Crab region, adjust for the different exposures and 
detector resolutions by normalization on the nearby background lines
at energies 1764 and 1779~keV,
and subtract this normalized ``off-source'' reference  
from the integrated spectrum measured from the
inner Galaxy, we obtain an excess signal which indicates the presence
of a celestial signal from \Al (Fig \ref{spec_gcde-crab}). 
The strong instrumental background lines do not perfectly subtract. 
The suppression by a factor \about~90, however,
would correspond to residuals from the instrumental feature at 1809~keV
of less than half of what we observe; also, residuals appear in the
wings of the instrumental lines, because 
our normalization does not trace time-variable gains or non-Gaussian
detector degradation. The feature at 1809~keV, however,
resembles the expected line more closely than background residuals:
we indicate the expected \Al signal with a Gaussian
at the \Al line energy and instrumental line width. Note that 
here we do not use SPI's imaging capabilities, so it is not 
surprising that a not very significant \Al detection is obtained
(see Figure \ref{spec_gcde-crab}).

Imaging analysis makes use of the detailed response of the SPI instrument
as is obtained from Monte Carlo simulations \citep{stur03}, including
the properties of the coded mask, the Ge camera, and all other 
material possibly interacting with incident gamma-rays. Calibrations
before INTEGRAL's launch \citep{atti03} and on Cyg-X1 and the Crab in the first
part of the mission have verified the validity of this response \citep{roqu03,stur03}.
Adopting an intensity distribution on the sky, one may thus predict
the measured event distribution from \Al for the observed exposure pattern.
Adding these to a suitable background model, one obtains a prediction
for the actually-measured dataset. A fitting procedure can then be used
to adjust intensity parameters of the \Al skymap and background components:
$$ D_{i,j,E}=I_E\cdot S \otimes \Re + a_{j,E,k} \cdot B_{E,k} + \aleph_{i,j,E}$$
Here $D$ are the measured data, $S$ the sky intensity distribution, $\Re$ the 
instrument response function, $B$ the background model, 
$\aleph$ the statistical noise, and indices are $i$
for detectors, $j$ for pointings, $E$ for energy, 
and $k$ for background model components.
From such fits we obtain intensity spectra $I_E$ of our
sky model as constrained by our measurement. 
Several implementations 
(e.g. {\it spidiffit/spi\_obs\_fit}) of 
this analysis concept have been developed \citep{stro03,knoe03um},
differing in the method of minimum searches and uncertainty estimates:
The fit determines the mean of the intensity parameter posteriors through a 
Markov-Chain Monte Carlo method marginalizing over the 
background parameters ({\it spidiffit}), or alternatively
minimizes the log-likelihood function through a Levenberg-Marquardt minimum search
to fit spectral intensities ({\it spi\_obs\_fit}). 
Background amplitudes
$a_{j,E,k}$ are fitted per pointing and energy from the measurement ({\it spidiffit}), 
or prescribed by an absolute model based on tracers of continuum and line 
background components ({\it spi\_obs\_fit}).
Error bars $\delta I_E$ are determined as standard deviation from the mean 
with the same Markov-Chain Monte Carlo method
\citep{stro03} for the large number of fitted parameters in 
 {\it spidiffit}, and  
 with an eigenvalue analysis of the error matrix
\citep{stro85,knoe03um} for the few fitted parameters in {\it spi\_obs\_fit}.

Estimates of the
systematic uncertainties have been added in quadrature to these statistical uncertainties
and are thus included in the error bars given in figures \ref{spec_diffit}--\ref{spec_cesr_87a}.
We fit different models ($k$) of background, such as continuum background
interpolated from adjacent energies, line backgrounds from scaled reference
observations off the source of interest, and scaling models using suitable
tracers of background such as the rate of saturated-signal events in the Ge
detectors; these are assumed to arise from cosmic-ray triggers which activate
spacecraft material and hence 
generate nuclear-line background \citep{jean03}. 
Suitable models $S$ of the \Al sky intensity distribution were adopted from
the COMPTEL \Al all-sky results \citep{plue01}, and from distributions of 
free electrons or warm dust as derived from 
COBE measurements \citep[see][]{knoe_mod99}.

If the signal is sufficient, one may consider imaging analyses in 
narrow spectral bins, such as exercised for the case of the 511~keV
emission from the inner Galaxy \citep{knoe03}; in our case, the signal
is too weak to reasonably constrain such analysis with its intrinsically
many free parameters.
For instrumental background
only, no or only modest correlation with a plausible \Al skymap is expected,
while the correlation should be significantly better within the \Al line
energy bins. However, systematic uncertainties in our background models lead
to distortions of such a correlation. It is the main task of current
analysis efforts to understand the level and possible spectral shapes
of such systematic uncertainties.  

\section{Results}

From spectral analysis through fitting of adopted models for the \Al 
skymap over an energy range around the \Al line, we obtain clear detections
of celestial \Al emission at the level of 5--7$\sigma$. The results
for the \Al flux, as well as details of the spectral signature, however, vary
significantly with parameters of the analysis, and thus indicate the levels
of uncertainty at this initial stage of the work; statistical uncertainties are
negligable, in comparison.

\begin{figure}[ht]
 \includegraphics[width=0.5\textwidth]{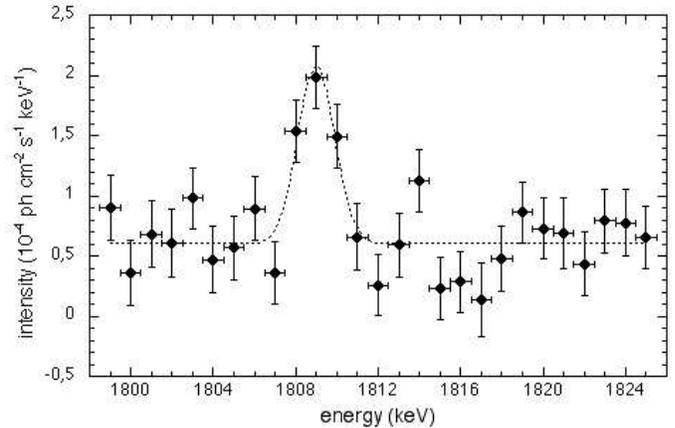}
   \caption{Imaging analysis result from fitting a sky intensity distribution
   as modelled from the COMPTEL \Al skymap to each energy bin. Background was 
   modelled from Crab observation detector ratios, and fitted in intensity
   to the actual measurement for each pointing, together with the sky signal.
    }
  \label{spec_diffit}
\end{figure}

\begin{figure}[ht]
 \includegraphics[width=0.5\textwidth]{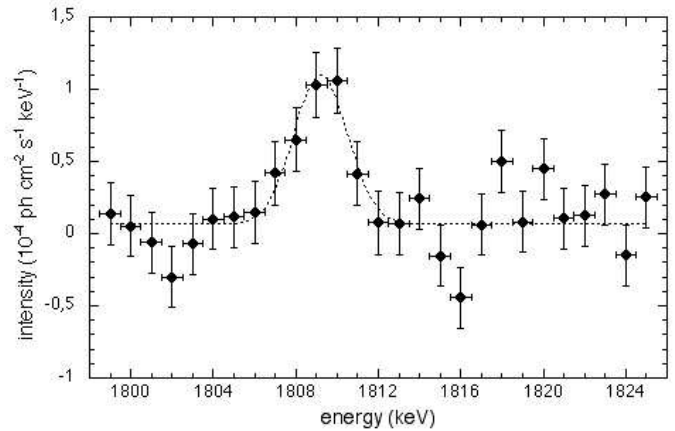}
   \caption{Imaging analysis result from fitting a sky intensity distribution
   as modelled from the COBE/DIRBE dust skymap (240$\mu$m) to each energy bin. 
   Background was modelled from adjacent energies for continuum, the Crab 
   observation detector ratios were used for the line, and scaled by the rate
   variations of saturated events in the detectors to model the line background.
}
  \label{spec_cesr_crab}
\end{figure}

\begin{figure}[ht]
 \includegraphics[width=0.5\textwidth]{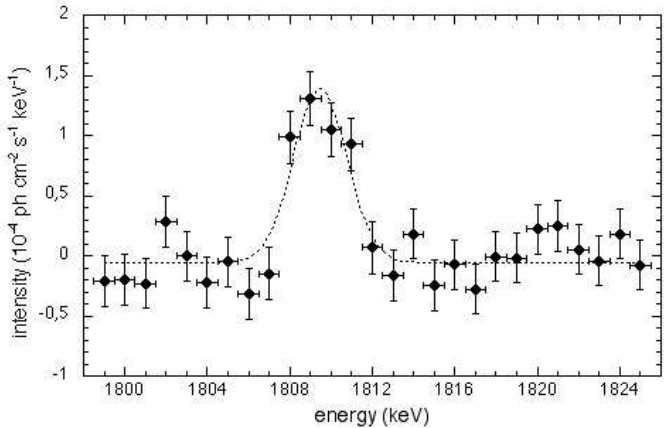}
    \caption{Same as Figure \ref{spec_cesr_crab}, 
     but for the exposures of the LMC/SN1987A
     used as background reference. 
	 }
  \label{spec_cesr_87a}
\end{figure}

\begin{table}\centering
\begin{tabular}{|l|l|l|}
\cline{1-3}
\vbox to1.88ex{\vspace{1pt}\vfil\hbox to15.80ex{\hfil \hfil}} & 
\vbox to1.88ex{\vspace{1pt}\vfil\hbox to12.80ex{\hfil FWHM [keV] \hfil}} & 
\vbox to1.88ex{\vspace{1pt}\vfil\hbox to18.80ex{\hfil I [$10^{-4}$ph~cm$^{-2}$s$^{-1}$] \hfil}} \\

\cline{1-3}
\vbox to1.88ex{\vspace{1pt}\vfil\hbox to15.80ex{\hfil uncertainty \hfil}} & 
\vbox to1.88ex{\vspace{1pt}\vfil\hbox to12.80ex{\hfil 0.7 \hfil}} & 
\vbox to1.88ex{\vspace{1pt}\vfil\hbox to18.80ex{\hfil 1.4 \hfil}} \\

\cline{1-3}
\vbox to1.88ex{\vspace{1pt}\vfil\hbox to15.80ex{\hfil fit value Fig. \ref{spec_diffit}\hfil}} & 
\vbox to1.88ex{\vspace{1pt}\vfil\hbox to12.80ex{\hfil 2.1 \hfil}} & 
\vbox to1.88ex{\vspace{1pt}\vfil\hbox to18.80ex{\hfil 3.3 \hfil}} \\

\cline{1-3}
\vbox to1.88ex{\vspace{1pt}\vfil\hbox to15.80ex{\hfil fit value Fig. \ref{spec_cesr_crab}\hfil}} & 
\vbox to1.88ex{\vspace{1pt}\vfil\hbox to12.80ex{\hfil 3.1 \hfil}} & 
\vbox to1.88ex{\vspace{1pt}\vfil\hbox to18.80ex{\hfil 3.3 \hfil}} \\

\cline{1-3}
\vbox to1.88ex{\vspace{1pt}\vfil\hbox to15.80ex{\hfil fit value Fig. \ref{spec_cesr_87a}\hfil}} & 
\vbox to1.88ex{\vspace{1pt}\vfil\hbox to12.80ex{\hfil 3.1 \hfil}} & 
\vbox to1.88ex{\vspace{1pt}\vfil\hbox to18.80ex{\hfil 4.7 \hfil}} \\

\cline{1-3}
\end{tabular}
\label{fitvalues}
\caption{The range of values fitted for several analysis scenarios illustrates the 
   range of systematic uncertainties.}
\end{table}

In Figure \ref{spec_diffit} we show a spectrum derived from all (single and
multiple) events with 
{\it spidiffit} using the COMPTEL
Maximum-Entropy map from 9 years of measurements as a model for the spatial
distribution of the sky emission \citep{plue01}. Given the rather modest 
spatial resolution of SPI, the particular choice of such distribution
is not critical, as long as the dynamic range of fluxes and spatial distribution
are approximately correct; any choice of good source tracers, such as the
warm dust or free electron distributions \citep[see e.g.][]{knoe_mod99}, 
produce very similar results.
Here we treat background by adopting the relative detector intensity ratios from
the Crab exposures, and adjust their intensity for each pointing
in the fit. 

In Figure \ref{spec_cesr_crab} we compare this to a spectrum derived 
from single events only (to avoid ME energy calibration issues), 
with {\it spi\_obs\_fit} and the 
COBE/DIRBE 240~$\mu$m dust map as sky model, modelling background for the continuum
from adjacent energy bins, and for the line component through taking detector
ratios from the Crab exposures and absolute intensity variations 
from the rate of saturated events in the Ge camera during the actual GCDE 
measurements.

Figure \ref{spec_cesr_87a} then illustrates how results depend on different
datasets for the background: here the exposures taken for SN1987A were used
as a reference to model detector ratios, in an otherwise analogous analysis
with  {\it spi\_obs\_fit}. 

Our fitted sky intensity values (see Table 1) 
from the inner $\pm$30$^\circ$ of the Galaxy
are  (3--5)$ \times 10^{-4}$~ph~cm$^{-2}$s$^{-1}$, and thus 
fall into the range suggested by previous observations: 
The general consensus for the inner-Galaxy brightness of \Al is
 \about~$4 \times 10^{-4}$~ph~cm$^{-2}$s$^{-1}$, considering all uncertainties involved
 \citep{pran96}; here 
  ``inner Galaxy'' means integrating over the central radian, 
 roughly $\pm 30^\circ$ in longitude about the Galactic Center.
 COMPTEL had measured a value somewhat lower value of 
 $2.8 \pm 0.15 \times 10^{-4}$~ph~cm$^{-2}$s$^{-1}$ \citep{ober97}, 
 with their background subtraction
 from high-latitude observations possibly suppressing large-scale diffuse flux
 components. On the other hand, RHESSI recently measured a rather high value
 of $5.7 \pm 0.54 \times 10^{-4}$~ph~cm$^{-2}$s$^{-1}$ \citep{smit03} from earth
 occultation analysis of their measurements pointed at the sun. 
 
 The different values for our line positions, all somewhat higher than
 the expected value of 1808.7~keV, may indicate systematics in our
 energy calibration, the impact of detector degradation over the time of 
 measurements, or some structured underlying background effect which
 may shift our signal upward by a few tenths of a keV. On the other hand, all
 line width results which we obtain are consistent with SPI's instrumental
 resolution of 3~keV (FWHM) and thus
 support RHESSI's recent finding \citep{smit03}
 that the broad line reported by GRIS \citep{naya96} probably cannot be confirmed.
 Work is in progress to refine our spectral treatment and background modelling, in order
 to be able to further quantify and substantiate this conclusion.

\begin{acknowledgements}
SPI has been completed under the responsibility and leadership of CNES.
We are grateful to ASI, CEA, CNES, DLR, ESA, INTA, NASA and OSTC for support.
\end{acknowledgements}


\begin{thebibliography}{}

\bibitem[Attie et al.(2003)]{atti03}
				 Attie, D., Cordier, B., Gros, M., {\it et al.} 2003, A\&A, this volume
\bibitem[Chen et~al.(1995)]{chen95}
 Chen, W., Gehrels, N., and Diehl, R. 1995, ApJ, {444}, L57

\bibitem[Chen et~al.(1997)]{chen97}
  Chen, W., Diehl, R., Gehrels, N., {\it et al.} 1997, ESA-SP 382, 105

\bibitem[Diehl et~al.(1995)]{dieh95}
  Diehl, R., Dupraz, C., Bennett, K.,  {\it et al.} 1995, A\&A, 298, 445

\bibitem[Gehrels \& Chen(1996)]{gehr96}
  Gehrels, N., and Chen, W. 1996, A\&AS, 120, 331

\bibitem[Hermsen \& Winkler(2002)]{herm02}
  Hermsen, W., and Winkler, C. 2002, The INTEGRAL Mission, in Proc. of
  the XXII Moriond Astroph. Meeting
 
 \bibitem[Jean et al.(2003)]{jean03} 
				 Jean, P., Vedrenne, G., Roques J.-P., {\it et al.} 2003, A\&A, this volume
 
\bibitem[Kn\"odlseder et al.(2003)]{knoe03} 
				 Kn\"odlseder, J., Lonjou, V., Jean, P., {\it et al.} 2003, A\&A, this volume

\bibitem[Kn\"odlseder(2003)]{knoe03um} 
  {Kn\"odlseder, J.} 2003, {\it spi\_obs\_fit} User Manual, to be found at 
  http://www.cesr.fr/$\sim$jurgen/isdc/index.html

\bibitem[Kn\"odlseder et~al.(1999)]{knoe_mod99}
  Kn\"odlseder, J., Bennett, K., Bloemen, H.,  {\it et al.} 1999, A\&A, {344}, 68
  
\bibitem[Kn\"odlseder(1999)]{knoed_phd99} 
  {Kn\"odlseder, J.} 1999, ApJ, {510}, {915} 
  
\bibitem[Kn\"odlseder et~al.(1999)]{knoed_img99} 
 Kn\"odlseder, J., Dixon, D., Bennett, K.,  {\it et al.} 1999, A\&A, { 345}, {813}

\bibitem[Mahoney et al.(1982)]{maho82}
  Mahoney, W.~A., Ling, J.C., Jacobson A.S., Lingenfelter R.E. 1982, ApJ, 262, 742

\bibitem[Mahoney et al.(1984)]{maho84}
  Mahoney, W.~A., Ling, J.C., Wheaton W.A., Lingenfelter R.E. 1984, ApJ, 286, 578

				 
\bibitem[Naya et~al.(1996)]{naya96}
  Naya, J.~E., Barthelmy, S.D., Bartlett, L.M., {\it et al.} 1996, Nature, 384, 44

\bibitem[Oberlack(1997)]{ober97}
  Oberlack, U. 1997, Ph. D. Thesis, Technische Universit\"at M\"unchen

\bibitem[Pl\"uschke et~al.(2001)]{plue01}
  Pl\"uschke, S., Diehl, R., Sch\"{o}nfelder, V.,  {\it et al.} 2001, ESA SP-459, 55

\bibitem[Prantzos \& Diehl(1996)]{pran96}
  Prantzos, N., and Diehl, R. 1996, Phys. Rep., 267, 1

\bibitem[Roques et al.(2003)]{roqu03}
				 Roques, J.-P., Schanne, S., von Kienlin, A., {\it et al.} 2003, A\&A, this volume

\bibitem[Smith(2003)]{smit03}
				 Smith, D. 2003, ApJ, 589, L55

\bibitem[Strong (2003)]{stro03}
				 Strong, A.W.  2003, {\it spidiffit} User Manual, to be found at 
				 http://isdc.unige.de
				 
\bibitem[Strong(1985)]{stro85}	
	Strong, A.W. 1985, A\&A, 150, 273			 
\bibitem[Sturner et al.(2003)]{stur03}
				 Sturner, S.J., Shrader, C.R., Weidenspointner, G., {\it et al.} 2003, A\&A, this volume

\bibitem[Sturner \& Naya(1999)]{stur99}
  Sturner, S.~J., and Naya, J.~E. 1999, ApJ, 526, 200

\bibitem[Taylor \& Cordes(1993)]{tayl93} 
  Taylor, J.~H., and Cordes, J.~M. 1993, ApJ, 411, 674

\bibitem[Vedrenne et al.(2003)]{vedr03}
				 Vedrenne, G., Roques, J.-P., Sch\"{o}nfelder, V.,  {\it et al.} 2003, A\&A, this volume

\bibitem[Weidenspointner et al.(2003)]{weid03}
				 Weidenspointner, G., Kiener J., Gros M., {\it et al.} 2003, A\&A, this volume

\bibitem[Winkler et al.(2003)]{wink03}
				 Winkler, C., Courvoisier, T.C., DiCocco, G.,  {\it et al.} 2003, A\&A, this volume


\end{thebibliography}
\end{document}